\documentclass[12pt]{article}
\usepackage{amssymb}

\usepackage{amsmath}
\usepackage{setspace}
\usepackage{footmisc}
\usepackage{amsmath,amssymb}
\usepackage{graphicx}
\usepackage{authblk}
\usepackage{enumitem}
\usepackage[titletoc]{appendix}
\usepackage[top=3cm, bottom=3cm, left=3cm, right=3cm]{geometry}

\begin{document}

\title{Resolving the Induction Problem: Can We State with Complete Confidence via
Induction that the Sun Rises Forever?}
\author{Youngjo Lee}
\affil{Department of Statistics, Seoul National University}
\maketitle

\begin{abstract}
Induction is a form of reasoning that starts with a particular example and
generalizes to a rule, namely, a hypothesis. However, establishing the truth
of a hypothesis is problematic due to the potential occurrence of
conflicting events, also known as the induction problem. The sunrise
problem, first introduced by Laplace (1814), is a quintessential example of
the probability-based induction. In his solution, a zero probability is
always assigned to the hypothesis that the sun rises forever, regardless of
the number of observations made. This is a symptom of fundamental deficiency
of probability-based induction: A hypothesis can never be accepted via the
Bayes-Laplace approach. Alternative priors have been proposed to address
this issue, but they have failed to fully overcome the deficiency. We
investigate why this occurs and demonstrate that the confidence does not
exhibit such a deficiency, as it is not a probability and therefore does not
adhere to Bayes' rule. The confidence is neither a likelihood to allow not
only a reconciliation between epistemic and aleatory interpretations of
probability but also a resolution in agreement with the evidence by enabling
us to accept a hypothesis with complete confidence as a rational decision.
\end{abstract}

\section{Introduction}

Induction from experience is crucial for drawing valid inferences. Induction
is a form of reasoning that moves from particular examples to a rule, where
one infers a hypothesis (proposition, scientific theory) based on data.
Originally, the goal of science was to corroborate hypotheses such as `all
ravens are black' or to infer them from observational data. However, the
difficulty of deriving such inductive logic has been recognized as a problem
of induction since classical times. Pyrrhonian skeptic Sextus Empiricus
(1933) questioned the validity of inductive reasoning, `positing that a
universal rule could not be established from an incomplete set of particular
instances. Specifically, to establish a universal rule from particular
instances by means of induction, either all or some of the particulars can
be reviewed. If only some of the instances are reviewed, the induction may
not be definitive, as some of the instances omitted in the induction may
contravene the universal fact; however, reviewing all the instances may be
nearly impossible, as the instances are infinite and indefinite.' Hume
(1748) argued that inductive reasoning cannot be justified rationally
because it presupposes that the future will resemble the past. In response
to Hume's skepticism, Bayes (1763) proposed an inductive reasoning procedure
to address the question of whether one can rationally accept a hypothesis.
However, he might not have embraced the broad scope of application now known
as Bayesian epistemology, which was pioneered and popularized by Laplace
(1814) as inverse probability.

Probability has been used to {re}present uncertainties in inductive
reasoning. Kolmogorov (1933) presented formal {definition of} mathematical
probability as the key measure of uncertainty. There have been two main
strands in interpreting probability; epistemic and aleatory. In the
epistemic interpretation, the probability represents the degree of belief in
a hypothesis. Pawitan and Lee (2024) discussed four different epistemic
interpretations applied to hypothesis; subjective Bayesian, logical,
objective Bayesian and fiducial interpretation. Savage (1954) provided an
axiomatic basis for subjective probability as a degree of personal belief.
Keynes (1921) proposed a logical interpretation as a degree of rational
belief. Though his theory was simply too vague to have direct impact on
statistics and mathematics, it may influence logical probability of Carnap
(1950), the objective Bayesian probability (Jeffreys, 1961) and fiducial
probability (FP; Fisher, 1930). In aleatory interpretation, the probability
is applied to random events, representing either a long-run frequency (Von
Mises, 1928) or a propensity of the generating mechanism (Pawitan and Lee,
2024). For them, a non-repeatable single event such as the true status of a
hypothesis does not have a probability. Fisher (1922) introduced likelihood
as another measure of uncertainties for fixed unknowns. Recently, extended
likelihood has been extensively developed for prediction of random unknowns
(Lee and Nelder, 1996; Lee et al., 2017; Pawitan and Lee, 2024; Lee and Lee,
2024).

Royall (1997) paraphrased three type conclusions about hypotheses $G$ and $H$
via inductive reasoning:

C1: I believe $G$ to be true.

C2: I should act as if $G$ were true.

C3: The observation is evidence supporting $G$ over $H.$

\noindent Bayesian approach would be useful for C1, frequentist approach
such as hypothesis testing of Neyman and Pearson (1933) for C2, and
likelihood theories for C3 (Royall, 1997). Bayesian epistemic probabilities
have been applied to all types of hypotheses in various fields (Paulos,
2011). We review how Bayesian epistemic probabilities can be applied to draw
C1-C3. Fisher (1930) introduced the FP as an alternative to Bayesian
epistemic probability without assuming a prior. Throughout his life, he
believed that it is a true probability of `early writers' (Fisher, 1958)
such as Bayes (Barnard, 1987); Fisher used the term `probability' in the
same sense as used by Bayes (Barnard, 1987). However, it encountered
substantial criticisms, with ongoing controversy about its adherence to
probability (Edwards, 1977). According to Savage (1961), Fisher's FP is `a
bold attempt to make the Bayesian omelet without breaking the Bayesian egg'.
The FP has {recently} become of interest as a {form of} confidence. The
confidence is an extended likelihood (Pawitan and Lee, 2021), which is a
distinct concept from both probability and likelihood. Extended likelihood
has been introduced to deal with complex models with additional random
unknowns. Confidence concept gives external meaning outside the
confidence-interval context. The extended likelihood can access to the
frequentist probability, an objective certification not directly available
to the classical likelihood (Pawitan and Lee, 2024). For example, the
confidence interpretation can be allowed for predictive intervals of random
unknowns (Lee and Lee, 2024). In this paper, we argue that FP aligns with
the extended likelihood, and that Bayes might be considered to utilize an
extended likelihood. Efron (1998) noted, ``Maybe Fisher's biggest blunder
will become a big hit in the 21st century!'' In this paper, we demonstrate
alternative ways to draw C1-C3 via likelihood-based inductive reasoning. The
two distinct epistemic and aleatory interpretations of probability can be
reconciled through the confidence. We emphasize its synergy rather than
conflict between epistemic and aleatory interpretations.

\section{Epistemic Probability for Inductive Logic}

Current inductive logic is based on the idea that the probability represents
a logical relation between the hypothesis and the relevant observations. Let 
$G$ be a hypothesis, such as `all ravens are black' or `the sun rises
forever' and $E$ be a particular observation or event such as `the raven in
front of me is black' or `the sun will rise tomorrow'. The deductive logic
that $G$ implies $E$ can be represented by the conditional probability 
\begin{equation*}
P(E|G)=1\text{ \ and \ }P(\text{not }E|G)=0
\end{equation*}
and its contrapositive logic, that not $E$ implies not $G,$ by 
\begin{equation*}
P(\text{not }G|\text{not }E)=1\text{ \ and \ }P(G|\text{not }E)=0.
\end{equation*}
The probability can be quantified as a number between 0 and 1, where 0
indicates impossibility (false), and 1 indicates certainty (true). Thus,
deductive reasoning allows us to predict a particular event and to falsify a
hypothesis by a conflicting observation. A single observation of a non-black
raven can certainly falsify the hypothesis.

Inductive reasoning has been studied based on the Bayes (1763) rule, defined
by the conditional probability 
\begin{equation}
P(G|E)=\frac{P(E|G)P(G)}{P(E)}=\frac{P(E,G)}{P(E)},  \label{eq:bayes}
\end{equation}
where $P(G)$ is called a prior of hypothesis $G$, $P(E|G)$ is a likelihood
and $P(G|E)$ is a posterior of $G$. This Bayes rule is implied by coherency
axiom in subjective probability theory (Savage, 1972), whereas by a
definition of the conditional probability in Kolmogorov system. Under
current system, probability obeys the Bayes rule. Provided that the
denominator is not zero, this leads to 
\begin{equation*}
1\geq P(G|E)=\frac{P(E|G)P(G)}{P(E)}=\frac{P(G)}{P(E)}\geq P(G)\geq 0.
\end{equation*}
Updating the degree of belief of $G$ from $P(G)$ to $P(G|E)$ via the Bayes
rule is called `learning from the data' in Bayesian epistemology. It is
prohibited to assert that $P(G)=0$ or $1$ because learning from the data is
impossible, as 
\begin{equation*}
P(G|E)=P(G).
\end{equation*}
A subjective Bayesian would consider people rational if they maximize the
expected utility of betting. De Finetti (1972) insisted that the betting
quotient $P(G)$ represents a degree of belief for computing the expected
utility. Assigning $P(G)=1$ means that one is willing to bet everything on
the hypothesis $G$. Since we are rarely willing to bet everything on a
hypothesis, we should avoid assigning $P(G)=1$ (Olsson, 2018). The `Bayesian
challenge' means that belief $P(G)=1$ is not relevant to rational
decision-making. Let $I(\cdot )$ be the indicator function such that $I(G)=1$
if hypothesis $G$ is true and $=0$ otherwise. Epistemic probability $P(G)=1$
or $0$ is {nevertheless} equivalent to $I(G)=1$ or $0$, respectively.
Prohibiting a complete belief in a hypothesis $P(G)=1$ would be compulsory
because it implies $I(G)=1$, which can be never known. This would be what
the Bayesian challenge avoids.

\section{Bayesian Solutions to the Sunrise Problem}

By using the sunrise problem, Laplace (1814) demonstrated how to update
degrees of belief (epistemic probability) on a hypothesis based on the data.
Let $\theta $ be the long-run frequency of sunrises, i.e., the sun rises on
100$\times \theta $\% of days. Under the Bernoulli model, the hypothesis $G$
that the sun rises forever is equivalent to hypothesis $\theta =1.$ Based on
the finite observations ($E$) until the present day, can we attain $%
P(G|E)=1? $

Prior to the knowledge of any sunrise, let us suppose that one is completely
ignorant of the value of $\theta $. Laplace (1814) represented this prior
ignorance due to the principle of insufficient reason by means of a uniform
prior $P_{0}(\theta )=1$ on $\theta \in (0,1).$ This uniform prior was also
proposed by Bayes (1763). Here, the probability of there being a sunrise
tomorrow is $\theta $. However, we do not know the true value of $\theta $.
Let $T_{n}$ be the number of sunrises in the last $n$ days. We are provided
with the observed data that the sun has risen every day on record ($T_{n}=n)$%
. Laplace, based on a young-earth creationist reading of the Bible, inferred
the number of days according to the belief that the universe was created
approximately 6000 years ago. The Bayes--Laplace rule gives the posterior
density, 
\begin{equation}
f(\theta |T_{n}=n)=\frac{P_{0}(\theta )\theta ^{n}}{\int_{0}^{1}P_{0}(\theta
)\theta ^{n}d\theta }=(n+1)\theta ^{n}.
\end{equation}
Consequently, the probability statements for $\theta $ can be established
from this posterior. In the Appendix, given $n=6000\times 365=2,190,000$
days of consecutive sunrises, we have 
\begin{equation*}
P(E|T_{n}=n)=\int \theta f(\theta |T_{n}=n)d\theta =\frac{n+1}{n+2}=\frac{%
2190001}{2190002}=0.9999995.
\end{equation*}
The probability of this particular event, that is, the sun rising the next
day, eventually becomes one as the number of observations increases.
However, this aspect is not sufficient to accept the hypothesis $G$ that the
sun rises forever. As shown in the Appendix, Broad (1918) showed that for
any $t=0,1,\cdots ,n$ and $n=1,2,\cdots ,$ 
\begin{equation*}
P(G|T_{n}=t)=0.
\end{equation*}
Thus, the Bayes-Laplace solution cannot be used for the significant test
with a null hypothesis $G:\theta =1,$ because it has a false-negative error
of 1. Jaynes (2003) argued that a beta prior, $Beta(\alpha ,\beta )$ with $%
\alpha >0$ and $\beta >0,$ describes the state of knowledge that $\alpha $
successes and $\beta $ failures were observed prior to the experiment. The
Bayes--Laplace uniform (equivalent to $Beta(1,1))$ prior means that one
success and one failure were observed \textit{a priori}. Having such a
uniform prior indicates that the experiment is a true binary in terms of
physical possibility $0<\theta <1$, leading to the parameter space $\Theta
=(0,1)$. Thus, it cannot be an ignorant prior. This explains why one cannot
accept $G$ ($P(G|E)=P(\theta =1|E)=0$) by using the Bayes rule (\ref
{eq:bayes}), as Bayes-Laplace's $Beta(1,1)$ prior assumes $P(G)=0$ \textit{a
priori}, i.e., $\theta =1\notin \Theta =(0,1)$, meaning $\theta =1$ is
outside of the parameter space. Even with an experiment with only successes
after tremendously large number of trials, there is no way to accept $G:$ $%
\theta =1$ unless the prior {of observing one failure} is discarded. There
is no way to attach even a moderate probability to $G$. Thus, the
Bayes-Laplace solution under a $Beta(1,1)$ prior cannot overcome the degree
of skepticism raised by Hume (1748). Thus, Broad {(1918)} stated that
`induction is the glory of science but the scandal of philosophy'. If there
is no way to accept any scientific hypothesis via data, it is also the
scandal of science. This is a peculiar phenomenon, called the probability
dilution when $G$ is true. Due to its simple nature, we use the sunrise
problem to explain how to overcome this fundamental deficiency of
probability-based induction.

Jeffreys's (1939) resolution was the use of another prior, which places a
point mass 1/2 on the null hypothesis $G:\theta =1$ and a point mass 1/2 on
the alternative hypothesis $G^{C}:\theta \thicksim Beta(1,1)$. Then, as
described in the Appendix, we have 
\begin{equation*}
P(E|T_{n}=n)=\frac{(n+1)(n+3)}{(n+2)^{2}}\text{ \ and \ }P(G|T_{n}=n)=\frac{%
n+1}{n+2}.
\end{equation*}
Senn (2009) considered Jeffreys's (1939) work to be `a touch of genius,
necessary to rescue the Laplacian formulation of induction' because it
overcomes probability dilution by allowing $P(G|T_{n}=n)>0$. As $n$
increases, $P(G|T_{n}=n)$ eventually goes to one under $G:\theta =1$. Under
this resolution, the probability dilution is alleviated, but it cannot
completely accept $G$ with finite observations because his prior places a
point mass of $1/2$ on $G^{C}$. Is there a way to accept $G$ with finite
observations? Jeffreys's great insight is to recognize the necessity of the
point mass at $\theta =1$ to avoid the probability dilution of $G$. Can we
allow a point mass without assuming any prior?

According to Hartmann and Sprenger (2010), there are three pillars of
Bayesian epistemology: (i) the Dutch book argument by Ramsey (1926) and de
Finetti (1972), (ii) the principal principle of Lewis (1980) and (iii)
Bayesian conditionalization. A logical device called the Dutch book has been
used to establish Bayesian subjective probability as an internally
consistent personal betting price for a single event. The personal degree of
belief about the hypothesis can be represented by betting odds. In de
Finetti's (1972) framework, betting odds should obey probability axioms to
avoid the Dutch book. In effect, his Dutch book argument assumes objective
utility (money) and a coherent betting strategy to arrive at subjective
probabilities. Von Neumman and Morgenstern (1947) assume that the aleatory
probability and axioms of rational preferences lead to subjective utility.
Savage (1954) combined the Dutch book argument and von Neumman and
Morgenstern's (1947) preferences in game theory to unify subjective
probability and subjective utility within a decision framework. His
axiomatic approach has great theoretical implications; prior probability can
be elected from subjective preferences and then updated with observations by
using the Bayes rule. Prior could reflect either ignorance about the
hypothesis in objective Bayesian school or one's fullest possible knowledge,
reflecting personal preference in subjective Bayesian school. Coherent
betting argument would be convincing only for the latter (Vovk, 1993).
Lewis's (1980) principal principle states that subjective probability must
be set as equal to objective probability if the latter exists. Logical,
objective Bayesian and fiducial probabilities have tried to identify an
objective epistemic probability. Bayesian conditionalization is the use of
the Bayes rule as learning from new observations. By setting the current
posterior $P(G|E)$ as the prior $P(G)$, the Bayes rule can update the
current degrees of belief $P(G|E^{\ast })$ in light of new observations $%
E^{\ast }$.

\section{Confidence and Extended Likelihood}

Neyman (1937) introduced the idea of confidence as aleatory coverage
probability of a confidence interval (CI) procedure. Recently, a surge of
renewed interest in confidence has arisen (Schweder and Hjort, 2016). Let $%
T_{n}$ be a random variable generated from a model $P(T_{n}|\theta )$ with
fixed unknown $\theta $ and $\theta _{0}$ be the true but unknown value of $%
\theta $. Suppose that we construct the $(1-\alpha )$ CI procedure $%
CI(T_{n}) $ for $\theta _{0}.$ We consider a binary latent variable 
\begin{equation*}
U=U(\theta _{0},T_{n})=I(\theta _{0}\in CI(T_{n})){,}
\end{equation*}
representing the true status of the CI procedure with 
\begin{equation*}
P(U=1)=P(\theta _{0}\in CI(T_{n}))=1-\alpha
\end{equation*}
being the aleatory coverage probability with a significant level $\alpha $.
Given an observation $T_{n}=t,$ we construct a $(1-\alpha )\times 100\%$
observed CI, namely $CI(t)$. Is there an objective epistemic concept to
state our sense of uncertainty in a given observed $CI(t)$ without assuming
a prior${?}$ For a given observation $T_{n}=t$, as an alternative to the
Bayesian posterior $P(\theta |t),$ Fisher (1930) defined a confidence
(fiducial) density 
\begin{equation}
c(\theta ;t)=\partial P(T_{n}^{\ast }\geq t|\theta )/\partial \theta ,
\label{eq:conf0}
\end{equation}
where $P(T_{n}\geq t|\theta )$ is the right side P-value and $T_{n}^{\ast }$
is another data set with the same distribution as $T_{n}$. We may call $%
c(\theta ;t)$ epistemic confidence among people with a common consensus
theory $P(T_{n}|\theta )$. Given $T_{n}=t,$ let 
\begin{equation*}
u=U(\theta _{0},t)=I(\theta _{0}\in CI(t){)}
\end{equation*}
be the realized as either one or zero but still unknown status of an
observed $CI(t)$ because $\theta _{0}$ is unknown. Pawitan and Lee (2021)
showed that the epistemic confidence of an observed CI can be computed as an
extended likelihood of $u,$ 
\begin{equation}
\ell _{e}(u=1;t)=C(\theta _{0}\in CI(t))=\int_{CI(t)}c(\theta ;t)d\theta
=1-\alpha ,  \label{eq:conf1}
\end{equation}
where $1-\alpha $ is the confidence of an observed $CI(t).$ Care is
necessary because it is not probability. Since 
\begin{equation*}
C(\theta _{0}\in CI(t))=1-\alpha \neq P(\theta _{0}\in CI(t))=I(\theta
_{0}\in CI(t))=u,
\end{equation*}
where $u$\ is either 0 or 1 but unknown, we use a distinct notation for the
confidence to avoid a probability-related controversies. For example, the
expected utility cannot be based on the confidence (Pawitan and Lee, 2017).
When $T_{n}$\ is continuous and sufficient, under appropriate conditions,
the confidence feature (Schweder and Hjort, 2016; Pawitan \textit{et al}.
2023) holds 
\begin{equation}
P(\theta _{0}\in CI(T_{n}))=C(\theta _{0}\in CI(t)),  \label{eq:confff}
\end{equation}
where the LHS is aleatory coverage probability of the CI procedure $%
CI(T_{n}) $ and the RHS is epistemic confidence of observed interval $CI(t).$
Likelihood procedures have been criticized for providing exact inferences
only asymptotically. Bayesian procedures provide exact interval estimation
in finite samples but have been criticized for assuming unverifiable prior.
The confidence feature of an observed interval indicates that an exact
interval estimation can be established for finite samples, but in a
frequentist sense, by maintaining the aleatory coverage probability of CI
procedure.

In the sunrise problem, the next section shows 
\begin{equation}
c(\theta ;t)=I(t=n)\mathbb{D}(\theta )+I(t<n)c_{+}(\theta ;t),
\label{eq:conff}
\end{equation}
where $\mathbb{D}(\theta )$ denotes the Dirac delta function to assign a
point mass one at $\theta =1$ and $c_{+}(\theta ;t)$ is the $Beta(t+1,n-t)$
density. Wilkinson (1977) called the point mass of confidence at $\theta =1$
a paradoxical unassigned probability because he presumed $\Theta =(0,1)$.
The uniform prior leads to posteriors without a point mass, so they always
allow two-sided credible interval $I(t)=(a(t),b(t))$ with $0<a(t)<b(t)<1,$
satisfying for all $0<\alpha <1$ and $t=0,\cdots ,n$ 
\begin{equation*}
P(\theta \in I(t)|T_{n}=t)=1-\alpha .
\end{equation*}
This is reasonable if $\Theta =(0,1).$ However, in this paper, we have $%
\Theta =(0,1]$ to allow a hypothesis $G:\theta =1.$ When $\theta _{0}=1\in
\Theta =(0,1]$, the coverage probability of two-sided credible interval
becomes $P(\theta _{0}\in I(t))=0$, because $\theta _{0}=1\not\in
I(t)=(a(t),b(t))$ for all $t$. This explains a probability dilution at $%
\theta _{0}=1$. When $\Theta =(0,1]$, the point mass at $\theta =1$ in $%
c(\theta ;t)$ is no longer an unassigned probability. Jeffreys's (1939)
prior puts a point mass of 1/2 at $\theta =1$, independent of the data $%
T_{n}=t.$ However, confidence has a data-dependent point mass of one to give
the complete confidence $C(\theta =1;n)=C(G;n)=1$ if $T_{n}=n$ and the null
confidence $C(G;t)=0$ if $T_{n}=t<n.$

Let $CI(t)$ be a two-sided observed CI if available. In the sunrise problem, 
$T_{n}$ is discrete, so the confidence feature (\ref{eq:confff}) holds
asymptotically. Figure 1 is a plot of the aleatory coverage probability $%
P(\theta _{0}\in CI(T_{n}))$ of the equal-tailed two-sided 95\% CI
procedures, against epistemic confidence $\ell _{e}(u=1;t)=C(\theta _{0}\in
CI(t))=1-\alpha =0.95$ of the hypothesis $\theta _{0}\in CI(t)$ for $n=10,$ $%
50$ and $500$ when $\Theta =(0,1].$ When $T_{n}=t<n,$ the confidence density 
$Beta(t+1,n-t)$ can always produce the two-sided CI for $\theta _{0}.$ When $%
\theta _{0}<1$, $T_{n}=t<n$ for some $n.$ When $\theta _{0}=1,$ $T_{n}=n$,
which leads to an one-sided oracle 100\% CI $\{1\}$. Because of a point mass
at $\theta =1$ the coverage probability of the CI is one when $\theta
_{0}=1. $ However, the complete confidence $C(G;n)=1$ does not necessarily
mean that $G:\theta =1$ is true. For example, when $\theta _{0}=0.99$ even
though $G$ is false$,$ $C(\theta =1;n)=C(G;n)=1$ with a probability $%
P(T_{n}=n|\theta =\theta _{0}),$ which can be quite large if $n$ is small.
Consequently, from Figure 1, we can see that the coverage probability at $%
\theta _{0}=0.99$ can be quite lower than the stated level 0.95 in small
samples. However, consistency means that the stated level 0.95 will be
achieved at $\theta _{0}=0.99$ eventually as $n$ increases. In summary,
Figure shows that the epistemic confidence $C(\theta _{0}\in CI(t))$ of an
observed $CI(t)$ is a consistent estimator of the aleatory coverage
probability $P(\theta _{0}\in CI(T_{n}))$ of the CI procedure $CI(T_{n})$
for all $\theta _{0}\in (0,1)$, i.e., $\ell _{e}(u=1;t)=1-\alpha \rightarrow
P(\theta _{0}\in CI(T_{n}))$ as $n\rightarrow \infty $. When $\theta _{0}=1,$
it is an oracle estimator $\ell _{e}(u=1;n)=$ $C(\theta _{0}\in
CI(T_{n}=n))=C(G;n)=1$ as if $\theta _{0}=1$ is known in advance$.$ Due to
its consistency, we can assure of our confidence as the number of
observations grows.

\begin{figure*}[tbp]
\centering
\includegraphics[width=\textwidth]{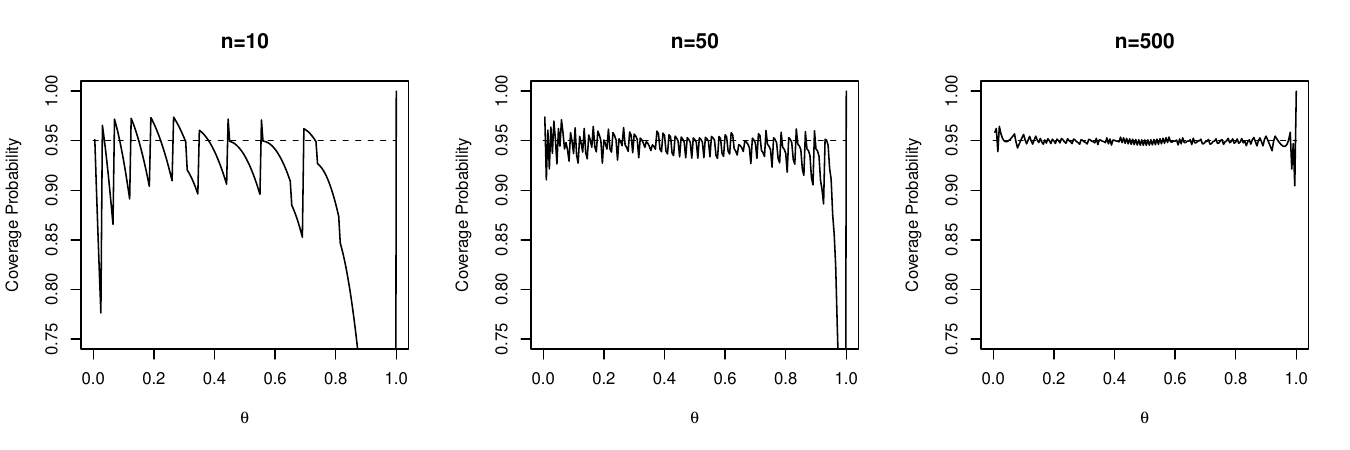}
\caption{The coverage probability against the 95\% confidence level at $%
n=10,50,500.$}
\end{figure*}

\section{Confidence Resolution}

In this section, we study how confidence can avoid a fundamental deficiency
of probability-based induction, namely the probability dilution. Popper
(1959) was a follower of the aleatory interpretation of probability.
However, for him, the main drawback of the aleatory view was its failure to
provide objective epistemic probabilities for single events. What Popper
wanted, I believe, was the confidence. Let us consider a single event for a
hypothesis $G$ 
\begin{equation*}
U=I(G).
\end{equation*}
Hypothesis testing is a prediction problem of discrete latent variable such
as $U=I(G)=0$ or $1$ (Lee and Bj{\o }rnstad, 2013). Given evidence $T_{n}=t,$
let $u$ be the realized but unknown status of the hypothesis $G$. Then, we
can compute the confidence (extended likelihood) for a hypothesis $G$ as 
\begin{equation}
\ell _{e}(u=1;t)=C(G;t).  \label{eq:confidence}
\end{equation}

Popper (1959) saw falsifiability as a criterion for scientific hypothesis;
if a hypothesis is falsifiable, it is scientific, and if not, then it is
unscientific. The hypothesis $G:\theta =1$ is then Popper scientific because
it can be falsified if a conflicting observation, i.e., one day of no
sunrise, occurs. Jeffreys's resolution corroborates Popper's theory that a
hypothesis is never accepted: Given $T_{n}=n,$ $P(G|T_{n}=n)<1$ for all $n$
and set $P(G|T_{n}=n)$ as a new $P(G).$ If a new conflicting evidence $%
X_{n+1}=0$ arrives we have a complete falsification $P(G|X_{n+1}=0)=0.$
Suppose now that $P(G|T_{n}=n)=1$ to set $P(G)=1,$ then 
\begin{align*}
P(G|X_{n+1}=0)
&=\frac{P(X_{n+1}=0|G)}{P(X_{n+1}=0|G)P(G)+P(X_{n+1}=0|\text{not }G)P(\text{not }G)} \\
&=\frac{0}{0}\text{ (undefined).}
\end{align*}
Thus, conflicting evidence cannot falsify the hypothesis with a prior $%
P(G)=1 $. To corroborate Popper's tentative acceptance theory, the prior and
posterior should never be set as one. Bayesian challenge is again requested
to support Popper's theory. Thus, probability cannot allow a paradigm shift
(Kuhn, 2012) (switching from $P(G)=1$ to $P(G)=0$) to accept a new
scientific theory. If probability is used for hypotheses, $P(G)$ is indeed
either $0$ or $1$ but unknown$.$ Popper's tentative acceptance theory
prohibits $P(G)=1$, so only $P(G)=0$ is allowed. Popper (1959, Appendix vii)
\ explained why he believed that $P(G)=0$ and used it as argument against
the probability-based induction because there is nothing to learn from the
data with $P(G)=0$. Indeed, he believed $I(G)=0$ \textit{a priori}. This
explains why the probability may not be adequate; he may object the use of
the Bayes rule for inductive reasoning.

Given $T_{n}=t$, the confidence density (Schweder and Hjort, 2016) can be
written as 
\begin{equation}
c(\theta ;t)\varpropto c_{0}(\theta ;t)L(\theta ;t),  \label{eq:confi}
\end{equation}
where $c_{0}(\theta ;t)$ is the implied prior, analogous to the Bayesian
prior and $L(\theta ;t)=P(T_{n}=t|\theta )$ is the Fisher (1921) likelihood.
Thus, the confidence density $c(\theta ;t)$ can be viewed as a Bayesian
posterior under an implied prior $c_{0}(\theta ;t).$ The Jeffreys invariant
prior has been advocated to improve the Bayes--Laplace uniform ($Beta(1,1))$%
\ prior, leading to Bernardo's (1979) reference prior for multi-parameter
cases. In single parameter cases, the invariant (equivalent to reference)
prior gives a asymptotically close posterior to $c(\theta ;t)$ (Lindley,
1958; Welch and Peers 1963). These priors have been developed as robust
priors in objective Bayesian school. In sunrise problem, the Jeffreys
(reference) prior is $Beta(1/2,1/2)$, but as shown in the Appendix, any $%
Beta(\alpha ,\beta )$ prior with $\alpha >0$ and $\beta >0$ cannot overcome
probability dilution; $P(G|T_{n}=n)=0$ for all $n.$ However, this problem
has been often ignored because it occurs only at a point $\theta =1$. In
Stein's (1959) high-dimensional problem, the use of uniform prior can lead
to a probability dilution in the entire parameter space. This causes that
the credible interval based on the uniform prior cannot maintain the stated
level of confidence. However, the reference prior cannot overcome the
probability dilution neither, whereas the confidence overcomes probability
dilution to allow a CI, maintaining the stated level of coverage
probabilities for all parameter values. The point mass of confidence (\ref
{eq:conff}) at boundary prevents such a probability dilution (Lee and Lee,
2023).

In Appendix, we show that $c_{0}(\theta ;t)=1/(1-\theta )$ in the sunrise
problem. Interpretation of the implied prior $c_{0}(\theta ;t)=1/(1-\theta )$
as $Beta(1,0),$ means that only one success is observed prior to the
experiment. This leaves room for a possibility of an experiment with
successes only, i.e., $\theta =1$. Consider a significant test with a null
hypothesis $G.$ Given $T_{n}=n,$ confidence allows us to accept $G$ with
complete confidence 
\begin{equation*}
C(G;n)=\ell _{e}(u=1;n)=1.
\end{equation*}
As a measure of consistency with the null hypothesis $G,$ $P(T_{n}<n|G)=0$
gives the actual level of significance attained by the data (Cox, 1958).
This explains why it accepts $G$ with complete confidence, $%
C(G;T_{1}=1)=P(T_{1}=1|G)=1,$ even with a single corroborating observation$.$
However, the confidence is not a probability but an extended likelihood.
Many controversies of confidence have stemmed from its confusion as a
probability (Pawitan and Lee, 2017; 2021; 2022; 2024). Even though we have
the complete confidence $C(G;n)=1$ for given $T_{n}=n,$ in contrast to the
Bayes rule, a single new conflicting observation $X_{n+1}=0$ can alter
complete confidence $C(G;T_{n}=n)=1$ to null confidence $C(G;T_{n+1}=n)=0$.
This shows that the epistemic confidence allows a switch from 1 to 0.
Current inductive reasoning uses the Bayes rule (\ref{eq:bayes}), by
presuming that $X_{n+1}$ and $G$ have a joint probability. In particular, it
implies that $G$ has an unverifiable prior probability $P(G)$. However, the
degree of confidence on the hypothesis $G$ is based on the whole data,
without using the conditional probability $P(G|X_{n+1})$. Consequently, the
integrated confidence is no longer a confidence; it may not eliminate
nuisance parameters by integration as Bayesian does. Therefore, the
marginalization paradox (Dawid et al., 1972) does not apply to confidence.
Probability allows marginalization and computation of expected utility. The
confidence $C(G;t)=1$ is in fact a simply consistent estimator of $I(G)=1$
based on an observation $T_{n}=t$, whereas the probability $P(G)=1$ is
equivalent to $I(G)=1.$

\section{Support interval Versus Confidence Interval}

Consistent with Carnap's (1950) corroboration theory, the Bayesian support
interval can be constructed based on the change from the prior to posterior
(PTP; Wagenmakers \textit{et al}. 2022), 
\begin{equation}
PTP(\theta ;t)=\frac{f(\theta |t)}{f(\theta )}=\frac{P(t|\theta )}{P(t)}%
\varpropto P(t|\theta )=L(\theta ;t),  \label{eq:support}
\end{equation}
where $P(t)=\int_{\Theta }P(t|\theta )f(\theta )d\theta $ and $L(\theta ;t)$
is the Fisher likelihood.\ The likelihood is a relative concept since it is
not a probability of $\theta ;$ $\int_{\Theta }L(\theta ;t)d\theta \neq 1.$
The Neyman-Pearson test is based on the likelihood ratio (LR) 
\begin{equation*}
LR(\theta _{1},\theta _{2};t)=\frac{P(T_{n}=t|\theta =\theta _{1})}{%
P(T_{n}=t|\theta =\theta _{2})}=\frac{L(\theta _{1};t)}{L(\theta _{2};t)}.
\end{equation*}
Thus, likelihood provides an immediate objective tool to compare hypotheses; 
$\theta _{1}$is preferred over $\theta _{2}$ if $LR(\theta _{1},\theta
_{2};t)>1.$\ However, traditional likelihood inference requires a sort of
probability-based calibration: see examples in Pawitan and Lee (2024),
showing how likelihood value alone can be misleading. Because $L(\theta
_{1};t)/L(\theta _{2};t)=PTP(\theta _{1};t)/PTP(\theta _{2};t),$ all the
evidence about $\theta $ in the data is in the PTP as well as in the
likelihood, (Birnbaum, 1962):$\ PTP(\theta ;t)$ can give maximum likelihood
estimation and Neyman-Pearson test (Wagenmakers \textit{et al}., 2022).
Since the likelihood $L(\theta ;t)$ is proportional to the posterior under
the flat prior $f(\theta )=1,$ the likelihood-based support interval can be
constructed based on the normalized likelihood 
\begin{equation*}
L_{n}(\theta ;t)=\frac{L(\theta ;t)}{\int_{\Theta }L(\theta ;t)d\theta }%
\varpropto L(\theta ;t).
\end{equation*}
Thus, the likelihood-based support interval can be interpreted as the
credible interval under the flat prior on $\theta $.

The Bayes rule (conditional probability) would be meaningless unless a joint
probability of $\theta $ and $y$\ is defined. Instead of assuming a prior on 
$\theta $ in (\ref{eq:confi}), in sunrise problem, let $c_{0}(\theta
;t)=|\partial \eta /\partial \theta |=1/(1-\theta )$ be the Jacobian term
for a particular scale $\eta =\eta (\theta ;t)=-\log (1-\theta )$ to satisfy 
$c_{0}(\eta ;t)\varpropto 1.$ Then, the confidence density is a normalized
likelihiood of the scale $\eta $ 
\begin{equation*}
c(\eta ;t)=L_{n}(\eta ;t),
\end{equation*}
where $\theta =\theta (\eta )=1-e^{-\eta }$, 
\begin{equation*}
c(\theta ;t)=L_{n}(\theta ;t)|\partial \eta /\partial \theta |=c(\eta
;t)|\partial \eta /\partial \theta |
\end{equation*}
and $L_{n}(\eta ;t)$ is the normalized likelihood with respect to $\eta $
such that $\int_{0}^{\infty }L_{n}(\eta ;t)d\eta =1$. Normalized likelihoods
are not invariant with respect to non-linear transformation of $\theta $ due
to the Jacobian term$.$ The CI is the likelihood-based support interval
under the flat prior on a particular scale $\eta $, allowing the
confidence-calibration 
\begin{equation*}
C(A;t)\equiv \int_{A}c(\theta ;t)d\theta =\int_{A^{\ast }}L_{n}(\eta
;t)d\eta ,
\end{equation*}
where $A^{\ast }=\{\eta :\eta =\eta (\theta ;t)$ for $\theta \in A\}.$\
Thus, the CI has an objective alreatory interpretation, by satisfying the
confidence feature (\ref{eq:confff}). In Stein's (1959) example, the scale $%
\eta $ is depending upon the observed data (Lee and Lee, 2023).

The choice of prior may correspond to that of scale to define the
likelihood-based support interval. For example, in sunrise problem, the
Jeffreys invariant (reference) prior, $c_{0}(\theta ;t)=|\partial \eta
/\partial \theta |=\theta ^{-1/2}(1-\theta )^{-1/2},$ leads to the
posterior, which is a normalized likelihood in the scale $\eta =Beta(\theta
;1/2,1/2)=\int_{0}^{\theta }t^{-1/2}(1-t)^{-1/2}dt,$ where $Beta(\theta
;1/2,1/2)$ is the incomplete beta function with $Beta(1;1/2,1/2)=Beta$
(1/2,1/2). Thus, all the credible intervals could be viewed as
likelihood-based support intervals under the flat prior on some scales.
Fraser and McDunnough (1984) showed that the asymptotic normal distribution
of the maximum likelihood estimators and normalized likelihoods are
approximations to the confidence density $c(\theta ;t).$ Thus, all the
support intervals satisfy the confidence feature asymptotically. Choice of
scale is important for a good approximation in finite samples. For example,
the normalization transform of $\theta $ for the Wald interval based on
asymptotic normality and the Jeffreys (reference) prior lead to good
approximations to the confidence density (Welch and Peers, 1963).

A major contribution of Lee and Nelder (1996) is a specification of a scale
of random parameters to define the h-likelihood among extended likelihoods
for prediction of random unknowns. Now, consider Bayes's (1763) billiard
example. Suppose $n+1$ balls are randomly thrown onto a table of length 1,
and let $\theta $ be an unobserved position of the first ball. Then, we
observe whether the $n$ other balls come to rest to the left or right of the
first ball. Among the $n$ balls, $T_{n}=t$ balls are observed to be to the
left of the first ball. Bayes identified that $\theta $ is a scale of random
parameter, which follows a uniform distribution, $Beta(1,1)$. Using extended
likelihood, Lee and Lee (2024) showed that a predictive interval (equivalent
to credible interval) can be made for an unknown realized value $\theta _{0}$
of the random parameter $\theta,$ 
which satisfies the confidence feature
with a confidence interpretation (\ref{eq:confff}). 
Thus, the billiard
example can be accommodated by confidence approach. 

To have a confidence interval with full data evidence, Pawitan \textit{et al}%
. (2023) proposed conditioning on ancillary statistics: for more discussion,
see Pawitan and Lee (2024). Cox (1958) addressed necessities of nuisance
parameters for a better approximation to the truth and of robust models,
less sensitive to model assumptions. Robust heavy-tailed models can be build
by adding random effects in various dispersion parameters of the model (Lee 
\textit{et al}., 2017). Recently, Lee and Lee (2024) extended the confidence
density for fixed unknowns to models with additional random unknowns. These
extensions lead to multi-parameter models with nuisance parameters, in which
the normalized likelihoods can give asymptotically exact support intervals
in Wagenmakers \textit{et al.} (2022). It is challenging to extend the
current exact CI theory mainly for a single parameter to that for
multi-parameters.

\section{Bayes Factor Versus Extended likelihood Ratio}

In this section, we study Bayesian and likelihood procedures for
Neyman-Pearson type hypothesis testing. In the sunrise problem, the Jeffreys
resolution leads to the Bayes factor (BF) for the simple null $G:\theta =1$
versus a general alternative $G^{C}:\theta \backsim Beta(1,1)$. The PTP for $%
G^{C}$ becomes 
\begin{equation*}
PTP(G^{C};t)=\frac{P(G^{C}|t)}{P(G^{C})}=\frac{P(t|G^{C})}{P(t)}\varpropto
P(t|G^{C}),
\end{equation*}
which yields an extension of likelihood to a general hypothesis $G^{C}$.
This leads to the BF 
\begin{equation*}
BF(G,G^{C};t)=\frac{P(t|G)}{P(t|G^{C})}=\frac{P(G|t)}{P(G^{C}|t)}\frac{%
P(G^{C})}{P(G)},
\end{equation*}
which is an extension of the LR for the general hypotheses, $G:\theta =1$
versus $G^{C}:\theta \thicksim Beta(1,1).$ Thus, $BF(G,G^{C};t)\geq k$ means
that the data supports $G$ over $G^{C}$ by $k$ times. 
In the sunrise problem, 
\begin{equation*}
BF(G,G^{C};t)=\frac{P(T_{n}=t|\theta =1)}
{P(T_{n}=t|\theta \thicksim Beta(1,1))}
=n+1
\text{ \ if }t=n
\text{\ \ and \ }=0\text{ \ if }t<n,
\end{equation*}
and
\begin{equation*}
BF(G,G^{C};t)=\frac{P(T_{n}=t|\theta =1)}
{P(T_{n}=t|\theta \thicksim Beta(1,1))}
=0
\text{ \ if }t<n,
\end{equation*}
Thus, the BF favors $G$ over $G^{C}$ if $t=n,$ and it accepts $G^{C}$ if $%
t<n.$ Thus, the BF can be used for testing general hypotheses.

Using the extended likelihood ratio (ELR), Lee and Bj{\o }rnstad (2013)
obtained the optimal multiple test. Is there an extension of the LR to a
composite alternative hypothesis $G^{N}:\theta <1?$ We may use the ELR 
\begin{equation*}
ELR(G,G^{N};t)=\frac{\ell _{e}(u=1;t)}{\ell _{e}(u=0;t)}=\frac{C(G;t)}{%
C(G^{N};t)}=\frac{C(\{1\};t)}{C((0,1);t)}=\infty \text{ \ if }t=n
\end{equation*}
and
\begin{equation*}
REL=0\text{ \ if }t<n.
\end{equation*}
This ELR accept $G$ if $t=n$ and accept $G^{N}$ if $t<n$. Note that BF and
ELR have different alternatives. The Fisher (1921) likelihood $L(\theta
;t)=P(T_{n}=t|\theta )$ represents evidence about $\theta $\ in the observed
data $T_{n}=t.$ However, the P-value $P(T_{n}\geq t|\theta )$ and therefore
the confidence uses evidence in other possible data sets $T_{n}^{\ast }>t$
in (\ref{eq:conf0}), generated from the theory $P(T_{n}|\theta ).$ Thus, it
is not likelihood but the extended likelihood, leading to confidence-based
calibration, 
\begin{equation*}
C(G;t)+C(G^{N};t)=C(\{1\};t)+C((0,1);t)=G(\Theta ;t)=1.
\end{equation*}
Confidence uses additional information of data generating process to allow a
sort of probability-based calibration. Thus, the likelihood and extended
likelihood are very distinct concepts: see discussion between Lavine and Bj{%
\o }rnstad (2022) and Pawitan and Lee (2022).

In this paper, we view both `confidence' and Jeffreys's resolution as a
consistent estimator (or predictor) of unobservable variable $U=I(G)$, the
true status of a hypothesis $G$. Both are asymptotically equivalent, so we
compare them in finite samples. In hypothesis testing, the type 1 error is
the rejection of a true null $G$, while the type 2 error is the rejection of
a true alternative. In practice, these two errors are traded off against
each other, i.e., an effort to reduce one usually increases the other. Since
it is impossible to avoid both errors, it is natural to consider the amount
of risk one is willing to take in deciding whether to accept null or
alternative. When the null $G$\ is true, the confidence resolution allows an
oracle acceptance with a null type 1 error $(REL(G,G^{N};n)=\infty )$
whereas the Jeffreys resolution has a tentative acceptance $(\infty
>BF(G,G^{C};n)>1)$ with a positive type 1 error. Because alternative
hypotheses of two resolutions are different, the type 2 errors may not be
directly compared. From a decision-making perspective, the confidence
resolution would be preferred to Jeffreys's resolution due to null type 1
error in finite samples. The Laplace solution under a uniform prior has a
type 1 error of one. As an alternative to probability-based reasoning,
accepting a scientific hypothesis with complete confidence is legitimate,
not irrational, as proclaimed by the `Bayesian challenge'.

The use of Haldane's prior $Beta(0,0)$ assumes the parameter space $\Theta
=[0,1]$ with an oracle property at both $\theta _{0}=0$ and $1.$ It leads to
decision making for three hypotheses; $\theta =0,$ $0<\theta <1$ and $\theta
=1.$ The use of consonant belief has been proposed (Denoeux and Li, 2018) to
overcome probability dilution. However, without plausibility (Dempster,
1968), a belief function alone may not be suitable for valid hypothesis
testing, in which confidence is preferred (Lee and Lee, 2023).

\section{Complementary Roles of Confidence}

Epistemic probability may not be entirely epistemic because $P(G)=1$ or $0$
is equivalent to the objective status $I(G)=1$ or $0,$ respectively. Fisher
was against the aleatory interpretation of probability because the orthodox
frequentist view is emphatic that it can be applied not to a single event
such as $I(G)$ but to a repeatable procedure. Thus, the epistemic
interpretation is unavailable or even denied in the frequentist school.
However, an epistemic interpretation is compulsory in science because
scientists may not repeat the same experiment again (Fisher, 1958) as the
frequentist CI procedure. Given that probability cannot suitably represent
uncertainty about $I(G)$, the meaning of Fisher's confidence (FP) has been
puzzling over the last century. Because of the confidence feature, for all $%
\theta _{0}\in {\Theta }$ and for all $n$ 
\begin{equation}
\ell _{e}(u=1;t)\equiv C({\theta _{0}\in CI(t)})\equiv P(U=1)=P(\theta
_{0}\in CI(T_{n})),  \label{eq:relax}
\end{equation}
epistemic confidence of the observed CI can agree with the aleatory
frequency of a repeatable procedure. Likelihood procedures have been
criticized for providing exact inferences only asymptotically. Bayesian
procedures provide exact interval estimation in finite samples but have been
criticized for assuming unverifiable prior. Confidence can allow exact
interval estimation in the frequentist perspective. Thus, two distinct
epistemic and aleatory interpretations of probability are reconciled with
the confidence.

Confidence provides complementary alternatives for the three pillars of
Bayesian epistemology. According to the Savage-Ramsey-de Finetti Dutch book,
a betting strategy is coherent as long as it satisfies the probability laws.
However, their definition of probability is personal because it deals with
for between you and me. Pawitan \textit{et al}. (2023) discussed how to use
confidence as a betting price for an observed CI, protected from the Dutch
book and showed that it can play a role in objective probability in Lewis's
(198) principal principle by not allowing a relevant subset. Due to a
consensus among scientists on the scientific theory $P(T_{n}|\theta )$, it
leads to an inter-subjective (market) price for the hypothesis. Conceptual
separation of personal and consensus market prices allows for both an
epistemic and an objective aleatory meaning of confidence. It allows a
logical interpretation as a degree of rational belief based on a consensus
scientific theory: see Pawitan and Lee (2024) for more discussion. The Bayes
rule is not used to update the confidence, which is simply based on the full
data. According to Kaplan (1996), a person with $P(G)=1$ should be willing
to bet on $G$ at any cost: One should not prefer the status quo (doing
nothing) to a bet on $G$ in which $G$ being true yields nothing and $G$
being false results in a million dollars because the personal expected
utility of both actions is the same. However, this is irrational. Thus,
Bayesian prohibits the term `belief', namely, $P(G)=1$. In epistemic
probability, $P(G)=1$ is equivalent to $I(G)=1$, which does not allow a
leaning from conflict data. However, in confidence, $C(G;T_{n}=n)=1$ is an
estimator of unknown $I(G)$ with a risk (type 2 error). With confidence the
zero estimated expected utility of the bet on $G$ is susceptible to risk, so
the status quo is preferred for having no risk. The consistency of complete
confidence means that the risk of accepting $G$ vanishes eventually as $%
n\rightarrow \infty $, i.e., $C(G;n)\rightarrow I(G)$ as $n\rightarrow
\infty .$ Thus, an estimator $C(G;n+1)=1$ with $T_{n+1}=n+1$ is more
convincing than an estimator $C(G;n)=1$ with $T_{n}=n$ because there is a
less risk for the former. In confidence approach, given finite observations,
oracle acceptance of the null hypothesis is rational to have null type 1
error. Such an oracle acceptance can be altered to a complete rejection with
null type 2 error as a new conflicting observation arises. Thus, the use of
an estimated expected utility without accounting for risk in estimation
would be irrational. Economists have been aware of this problem in using an
expected utility for decision making; various alternatives have been
proposed, for example Kahneman and Tversky's (1979) prospect theory: see
also Pawitan and Lee (2024) for more discussion.

\section{Concluding Remarks}

Frank (1954) studied a variety of reasons for the acceptance of scientific
hypotheses. Among scientists, it is taken for granted that a scientific
hypothesis should be accepted if and only if it is truly true ($I(G)=1)$.
Frank indicated that the `simplicity (or beauty) of a hypothesis' and
`agreement with the common sense' are important for a hypothesis to become
the consensus scientific theory. In practice, being true means being in
`agreement with observations' that can be logically derived from the
hypothesis; this agreement can be established by the consistent oracle
estimation of the true status of hypothesis $I(G)$ via confidence $C(G;t)$.
Russell (1912) illustrated the induction problem: `Domestic animals expect
food when they see the person who usually feeds them. All these rather crude
expectations of uniformity are liable to be misleading. The man who has fed
the chicken every day throughout its life at last wrings its neck instead,
showing that more refined views as to the uniformity of nature would have
been useful to the chicken.' We agree with Russell's claim that personal
belief of the dead chicken $P(G)=1$ is wrong because in reality $I(G)=0$.
However, the rest of chickens who witness the incidence $E$ will switch
their inter-subjective confidence from $C(G)=1$ to $C(G;E)=0$. Before
encountering an incidence $E$, confidence $C(G)=1$ is not wrong, but is
simply a bad estimator of $I(G)$, which can be altered correctly to $C(G;E)=0
$.

We have seen that inductive logic can be based on the confidence. The
Laplace--Bayes probability-based solution has a probability dilution of $%
P(G|T_{n}=n)=0$ even when $G$ is true. This is caused by the prior
assumption of $P(G)=0$, equivalently $I(G)=0$. With Jeffreys's (1939)
resolution, $T_{n}=n$ corroborates the hypothesis $G$ positively by putting
a point mass 1/2 on $G$ but never accept $G$, $P(G|T_{n}=n)<1$ with a finite 
$n$, because he set $P(G)=0$ with a point mass $1/2$ \textit{a priori}.
Confidence is not a probability, hence it does not have a probability
dilution. The confidence resolution allows an oracle acceptance of $%
C(G;T_{n}=n)=1$ when $G$ is true. It is a consistent oracle estimator of $%
I(G)$. The confidence is neither a probability nor a likelihood, to allow
for switching from 1 to 0, by not obeying the Bayes rule. It also gives a
meaningful epistemic confidence interpretation for an observed CI by
maintaining the confidence feature. With the confidence, to accept a
hypothesis from particular instances without an unverifiable prior,
inductive reasoning does not require to review all the instances but to have
a consensus scientific theory $P(T_{n}|\theta )$ on the generation of
instances. Then, we infer $G$ based on agreement with observed data,
predicted from $G$. It is legitimate to predict the uniformity until it
fails to hold. The general relativity theory was accepted in 1919 by
observing light bending as predicted by relativity theory. This approach
allows a dynamic paradigm shift (Kuhn, 2012) by accepting a new general
relativity theory and rejecting the former Newtonian theory. Such an oracle
acceptance of a new relativity theory with complete confidence is rational
(C1). This way of the rational acceptance and falsification of hypotheses is
theoretically consistent decision making (C2) and agrees with evidence from
the data (C3).

\section*{Appendix A: Bayesian approach}

Laplace (1814) used the Bernoulli model for the sunrise problem. Let $%
X=(X_{1},\cdots ,X_{n})$ be independent and identically distributed
Bernoulli random variables with the success probability $\theta .$ Once we
have observed data $x=(x_{1},\cdots ,x_{n}),$ we have the likelihood 
\begin{equation*}
L(\theta ;x)=P(X=x|\theta )\varpropto P(T_{n}=t|\theta )\varpropto \theta
^{t}(1-\theta )^{n-t}=L(\theta ;t),
\end{equation*}
where $T_{n}=\sum X_{i}$ is a sufficient statistic and $t=\sum x_{i}$.
Laplace (1814) used a uniform prior $P_{0}(\theta )=1$ on $\theta \in (0,1).$
To find the logical conditional probability of $\theta $ given $T_{n}=t$,
the Bayes-Laplace rule can be used: The conditional probability density of $%
\theta $ given the data $T_{n}=t$ is called the posterior 
\begin{equation}
f(\theta |t)=\frac{P_{0}(\theta )L(\theta ;t)}{\int L(\theta ;t)P_{0}(\theta
)d\theta }=\frac{P_{0}(\theta )L(\theta ;t)}{P(T_{n}=t)},
\label{eq:posterior}
\end{equation}
where 
\begin{equation*}
P(T_{n}=t)=\int L(\theta ;t)P_{0}(\theta )d\theta =\int \theta ^{t}(1-\theta
)^{n-t}d\theta =Beta(t+1,n-t+1),
\end{equation*}
and $Beta(\cdot ,\cdot )$ is the beta function. Let $E$ be a particular
event that the sun rises tomorrow. Then, 
\begin{equation*}
P(T_{n}=n)=Beta(n+1,1)=\frac{1}{n+1},
\end{equation*}
to give 
\begin{align*}
P(E|\text{ The sun has risen }n\text{ consecutive days})&
=P(X_{n+1}=1|T_{n}=n) \\
& =P(T_{n+1}=n+1|T_{n}=n) \\
& =\frac{n+1}{n+2}
=\frac{2190001}{2190002}=0.9999995.
\end{align*}
This shows that 
\begin{equation*}
P(E|T_{n}=n)\rightarrow 1\text{ as }n\rightarrow \infty .
\end{equation*}
One observation increases the probability of the particular hypothesis $E$
from $\frac{n}{n+1}$ to $\frac{n+1}{n+2},$ so that the increment of
probability is $\frac{n+1}{n+2}-\frac{n}{n+1}=\frac{1}{(n+1)(n+2)}.$ Thus,
the probability of this particular hypothesis eventually becomes one.
However, this is not enough to ensure the hypothesis $G$ that the sun rises
forever holds (Senn, 2003). The probability that the sun rises in the next $%
m $ consecutive days, given the previous $n$ consecutive sunrises, is 
\begin{equation*}
P(X_{n+1}=1,...,X_{n+m}=1|T_{n}=n)=P(T_{n+m}=n+m|T_{n}=n)=\frac{n+1}{n+m+1}.
\end{equation*}
As long as $n$ is finite, the probability of hypothesis $G$ becomes zero
because for all $n>0$ 
\begin{equation*}
P(G|T_{n}=n)=\lim_{m\rightarrow \infty }P(T_{n+m}=n+m|T_{n}=n)=0.
\end{equation*}
Let us consider Jeffreys's prior, which places 1/2 of probability on $\theta
=1$ and puts a uniform prior on (0,1) with 1/2 probability. Then, 
\begin{equation*}
P(T_{n}=n)=\frac{1}{2}\left( 1+\frac{1}{n+1}\right)
\end{equation*}
to give 
\begin{align*}
P(E|T_{n}=n)=P(T_{n+1}=n+1|T_{n}=n)& =\frac{1+1/(n+2)}{1+1/(n+1)}=\frac{%
(n+1)(n+3)}{(n+2)^{2}}, \\
P(T_{n+m}=n+m|T_{n}=n)& =\frac{1+1/(n+m+1)}{1+1/(n+1)}=\frac{(n+1)(n+m+2)}{%
(n+2)(n+m+1)}.
\end{align*}
Thus, 
\begin{equation*}
P(G|T_{n}=n)=\frac{n+1}{n+2}\rightarrow 1\text{ as }n\rightarrow \infty .
\end{equation*}

\section*{Appendix B: Confidence approach}

Let us consider the right-side P-value 
\begin{equation*}
P(T_{n}^{\ast }\geq t|\theta )=\sum_{y=t}^{n}P(T_{n}^{\ast }=y|\theta
)=\sum_{y=t}^{n}\frac{n!}{y!(n-y)!}\theta ^{y}(1-\theta )^{n-y}=\frac{%
\int_{0}^{\theta }x^{t-1}(1-x)^{n-t}dx}{Beta(t,n-t+1)}.
\end{equation*}
This leads to $Beta(t,n-t+1)$ density as the confidence density 
\begin{equation*}
c(\theta ;t)=\frac{\theta ^{t-1}(1-\theta )^{n-t}}{Beta(t,n-t+1)}.
\end{equation*}
Here, the implied prior 
\begin{equation*}
c_{0}(\theta ;t)\varpropto c(\theta ;t)/L(\theta ;t)\varpropto \theta ^{-1},
\end{equation*}
namely, the $Beta(0,1)$ distribution, is improper as $\int_{0}^{1}\theta
^{-1}d\theta =\infty .$ Necessary computations of $Beta(0,1)$ can be
obtained as the limit of a proper $Beta(a,1)$ distribution for $a>0$. $%
Beta(a,1)$ leads to 
\begin{equation*}
P(T_{n}=n)=\int_{0}^{1}P(T_{n}=n|\theta )c_{0}(\theta ;n)d\theta =\frac{%
\Gamma (a+1)}{\Gamma (a)}\int_{0}^{1}\theta ^{n+a-1}d\theta =\frac{\Gamma
(a+1)}{(n+a)\Gamma (a)}
\end{equation*}
and 
\begin{align*}
P(X_{n+1}=1|T_{n}=n)& =\frac{P(T_{n+1}=n+1)}{P(T_{n}=n)}=\frac{n+a}{n+a+1},
\\
P(T_{n+m}=n+m|T_{n}=n)& =\frac{P(T_{n+m}=n+m)}{P(T_{n}=n)}=\frac{n+a}{n+m+a}.
\end{align*}
Thus, 
\begin{equation*}
\lim_{a\downarrow 0}P(T_{n+m}=n+m|T_{n}=n)=\frac{n}{n+m}
\end{equation*}
Given $n,$ $Beta(0,1)$ implied prior leads to 
\begin{equation*}
C(G;T_{n}=n)=\lim_{m\rightarrow \infty }P(T_{n+m}=n+m|T_{n}=n)=0.
\end{equation*}
This confidence cannot overcome the degree of skepticism yet.

Now, we apply the confidence to the transformed data by defining $Y_{i}=0$
if the sun rises on the $i$th day, where $P(Y_{i}=1|\theta ^{\ast })=\theta
^{\ast }=1-\theta $ and $\theta $ is the long-run frequency of sunrises.
Then, $Y_{i}=1-X_{i}$ and $\theta ^{\ast }=1$ are equivalent to $\theta =0$.
Let $T_{n}^{\ast }=\sum Y_{i}=n-T_{n}$. Then, the right-side P-value
function is 
\begin{equation*}
P(T_{n}^{\ast }\geq n-t|\theta ^{\ast })=P(T_{n}\leq t|\theta ^{\ast
})=1-P(T_{n}\geq t+1|\theta ^{\ast })
\end{equation*}
for $t=0,1,\cdots ,n-1$. For $t=n$, we have for any $\theta ^{\ast }\in
\lbrack 0,1]$ 
\begin{equation*}
P(T_{n}\leq n|\theta ^{\ast })=1
\end{equation*}
which gives point mass 1 at $\theta ^{\ast }=0$, equivalently $\theta =1$.
The confidence distribution for $t=n$ leads to 
\begin{align*}
P(T_{n}^{\ast }=0)& =P(T_{n}=n)=1, \\
P(T_{n+m}^{\ast }=0|T_{n}^{\ast }=0)& =P(T_{n+m}=n+m|T_{n}=n)=1,
\end{align*}
and 
\begin{equation*}
C(G;T_{n}=n)=\lim_{m\rightarrow \infty }P(T_{n+m}=n+m|T_{n}=n)=1.
\end{equation*}
Thus, we can say that the sun will rise forever with complete confidence.
Because 
\begin{equation*}
c_{0}(\theta ;t)=c_{0}^{\ast }(\theta ^{\ast };n-t)\propto \frac{c^{\ast
}(\theta ^{\ast };n-t)}{L(\theta ^{\ast };n-t)}\propto \frac{1}{1-\theta },
\end{equation*}
we can obtain the same result from the implied prior $Beta(1,0)\varpropto
(1-\theta )^{-1},$ The computations for $Beta(1,0)$ can be obtained as the
limit of $Beta(1,a)$ for $a>0$, which leads to 
\begin{align*}
P(T_{n}=n)& =\int_{0}^{1}P(T_{n}=n|\theta )c_{0}(\theta ;n)d\theta \\
& =\frac{\Gamma (a+1)}{\Gamma (a)\Gamma (1)}\int_{0}^{1}\theta ^{n}(1-\theta
)^{a-1}d\theta \\
& =\frac{\Gamma (a+1)}{\Gamma (a)\Gamma (1)}\frac{\Gamma (n+1)\Gamma (a)}{%
\Gamma (n+a+1)}\int_{0}^{1}\frac{\Gamma (n+a+1)}{\Gamma (n+1)\Gamma (a)}%
\theta ^{n}(1-\theta )^{a-1}d\theta \\
& =\frac{\Gamma (a+1)\Gamma (n+1)}{\Gamma (n+a+1)}\rightarrow 1
\end{align*}
as $a\downarrow 0.$ Thus, $C(G;T_{n}=n)=1.$ With a prior $Beta(\alpha ,\beta
)$ where $\alpha >0$ and $\beta >0$, 
\begin{equation*}
P(X_{n+1}=1|T_{n}=n)=\frac{Beta(n+1+\alpha ,\beta )}{Beta(n+\alpha ,\beta )}=%
\frac{n+\alpha }{n+\alpha +\beta }.
\end{equation*}
Furthermore, 
\begin{align*}
\log P(T_{n+m}=n+m|T_{n}=n)
&=\log \left( \prod_{i=0}^{m-1}\frac{n+i+\alpha }{n+i+\alpha +\beta }\right) \\
&=\sum_{i=0}^{m-1}\log \left( 1-\frac{\beta }{n+i+\alpha +\beta }\right) \\
&\leq -\beta \sum_{i=0}^{m-1}\frac{1}{n+i+\alpha+\beta }.
\end{align*}
to give 
\begin{equation*}
0\leq C(G;T_{n}=n)=\lim_{m\rightarrow \infty }P(T_{n+m}=n+m|T_{n}=n)\leq
\exp (-\infty )=0,
\end{equation*}
provided $\beta >0.$

\section*{References}

\begin{description}
\item  Barnard, G. A. (1987). R.A. Fisher-a true Bayesian? \textit{%
International Statistical Review}, \textbf{55}, 182-189.

\item  {Bayes, T. } (1763). An Essay Towards Solving a Problem in the
Doctrine of Chances. By the late Rev. Mr. Bayes, F. R. S. communicated by
Mr. Price, in a Letter to John Canton, A. M. F. R. S.. \textit{Philosophical
Transactions of the Royal Society of London}, \textbf{53}, 370-418.

\item  {Bernardo, J. M. } (1979). Reference Posterior Distributions for
Bayesian Inference. \textit{Journal of the Royal Statistical Society Series
B: Statistical Methodology}, \textbf{41}, 113-128.

\item  Birbaum, A. (1962). On the foundation of statistical inference. 
\textit{Journal of the American Statistical Association}, \textbf{57},
269-326.

\item  {Broad, C. D. } (1918). On the Relation Between Induction and
Probability (Part I.). \textit{Mind}, \textbf{27}, 389-404.

\item  {Cox, D. R. } (1958). Some problems connected with statistical
inference. \textit{The Annals of Mathematical Statistics}, \textbf{29},
357-372.

\item  {Dawid, A. P., Stone, M. and Zidek, J. V.} (1973). Marginalization
paradoxes in Bayesian and structural inference. \textit{Journal of the Royal
Statistical Society Series B: Statistical Methodology}, \textbf{35}, 189-233.

\item  {De Finetti, B. } (1972). \textit{Probability, Induction and
Statistics}. (London: J. Wiley.)

\item  {Dempster, A. P. } (1968). Upper and lower probabilities generated by
a random closed interval. \textit{The Annals of Mathematical Statistics}, 
\textbf{39}, 957-966.

\item  {Denoeux, T. and Li, S. } (2018). Frequency-calibrated belief
functions: Review and new insights. \textit{International Journal of
Approximate Reasoning}, \textbf{92}, 232-254.

\item  {Empiricus, S. } (1933). \textit{Outlines of Pyrrhonism}. (Translated
by R. G. Bury. London: W. Heinemann)

\item  {Fisher, R. A. } (1921). On the probable error' of a coefficient of
correlation deduced from a small sample. \textit{Metron}, \textbf{1}, 1-32.

\item  {Fisher, R. A. } (1930). Inverse Probability. \textit{Mathematical
Proceedings of the Cambridge Philosophical Society}, \textbf{26}, 528-535.

\item  {Fisher, R. A. } (1958). The Nature of Probability. \textit{%
Centennial Review of Arts and Science}, \textbf{2}, 261-274.

\item  {Frank, P. G. } (1954). The variety of reasons for the acceptance of
scientific theories. \textit{The Scientific Monthly}, \textbf{79}, 139-145.

\item  {Fraser, D. A. S. and McDunnough, P. } (1984). Further remarks on
asymptotic normality of likelihood and conditional analyses. \textit{%
Canadian Journal of Statistics}, \textbf{12}, 183-190.

\item  {Hartmann, S. and Sprenger, J. } (2010). Bayesian Epistemology. (In
D. Pritchard and S. Bernecker (Eds.), \textit{The Routledge Companion to
Epistemology}. (London: Routledge)

\item  {Hume, D. } (1748). \textit{An Enquiry Concerning Human Understanding}%
, P. Millican (Ed.). (Oxford: Oxford University Press)

\item  {Jaynes, E. T. } (2003). \textit{Probability Theory: The Logic of
Science}, G. L. Bretthorst (Ed.). (Cambridge: Cambridge University Press)

\item  {Jeffreys, H. } (1939). \textit{Theory of Probability}. (New York:
Oxford University Press)

\item  {Kahneman, D. and Tversky, A. } (1954). Prospect theory: Analysis of
decision under risk. \textit{Econometrica}, \textbf{47}, 263-292.

\item  {Kaplan, M. } (1996). \textit{Decision Theory as Philosophy}.
(Cambridge: Cambridge University Press)

\item  {Kolmogorov, A. N. } (1933). \textit{Foundations of Probability}.
(Translated by N. Morrison. New Yourk: Chelsea Publishing Company)

\item  {Kuhn, T. S. } (2012). \textit{The Structure of Scientific Revolutions%
}, 4th ed. with I. Hacking (intro). (University of Chicago Press)

\item  {Laplace, P. S. } (1814). \textit{A Philosophical Essay on
Probabilities}. (Translated by F. W. Truscott and F. L. Emory. New York:
Dover.)

\item  {Lavine, M., and Bj{\o }rnstad, J. F.} (2022). Comments on Confidence
as Likelihood by Pawitan and Lee in Statistical Science, November 2021. 
\textit{Statistical Science}, \textbf{37}, 625-627.

\item  {Lee, H. and Lee, Y. } (2023). Point mass in the confidence
distribution: Is it a drawback or an advantage?. \textit{arXiv preprint
arXiv:2310.09960}.

\item  {Lee, H. and Lee, Y. } (2024). On the Statistical Foundations of
H-likelihood for Unobserved Random Variables. \textit{arXiv preprint
arXiv:2310.09955}.

\item  {Lee, Y. and Bj{\o }rnstad, J. F. } (2013). Extended Likelihood
Approach to Large-scale Multiple Testing. \textit{Journal of the Royal
Statistical Society Series B: Statistical Methodology}, \textbf{75}, 553-575.

\item  {Lee, Y. and Nelder, J. A. } (1996). Hierarchical Generalized Linear
Models. \textit{Journal of the Royal Statistical Society Series B:
Statistical Methodology}, \textbf{58}, 619-678.

\item  {Lee, Y., Nelder, J. A. and Pawitan, Y. } (2017). \textit{Generalized
Linear Models with Random Effects: Unified Analysis via H-likelihood}, 2nd
ed. (Boca Raton: CRC Press)

\item  {Lewis, D. } (1980). A Subjectivist Guide to Objective Chance. (In W.
L. Harper, R. Stalnaker and G. Pearce (Eds.), \textit{Ifs}. Berkeley:
University of California Press)

\item  {Lindley, D. V.} (1958). Fiducial distributions and Bayes' theorem. 
\textit{Journal of the Royal Statistical Society Series B: Statistical
Methodology}, \textbf{20}, 102-107.

\item  {Meng, X. L. } (2009). Decoding the h-likelihood. \textit{Statistical
Science}, \textbf{24}, 280-293.

\item  {Neyman, J. } (1937). Outline of a theory of statistical estimation
based on the classical theory of probability. \textit{Philosophical
Transactions of the Royal Society of London. Series A, Mathematical and
Physical Sciences}, \textbf{236}, 333-380.

\item  {Neyman, J. } (1956). Note on an Article by Sir Ronald Fisher. 
\textit{Journal of the Royal Statistical Society Series B: Statistical
Methodology}, \textbf{18}, 288-294.

\item  {Neyman, J. and Pearson, E. S. } (1933). On the problem of the most
efficient tests of statistical hypotheses. \textit{Philosophical
Transactions of the Royal Society of London. Series A, Containing Papers of
a Mathematical or Physical Character}, \textbf{231}, 289-337.

\item  {Olsson, E. J. } (2018). Bayesian Epistemology. in S. O. Hansson and
V. F. Hendricks (eds), \textit{Introduction to Formal Philosophy}.

\item  {Paulos, J. A. } (2011). The Mathematics of Changing Your Mind. 
\textit{New York Times}, 5 August 2011.

\item  {Pawitan, Y., Lee, H. and Lee, Y. } (2023). Epistemic confidence in
the observed confidence interval. \textit{Scandinavian Journal of Statistics}%
, \textbf{50}, 1859-1883.

\item  {Pawitan, Y. and Lee, Y. } (2017). Wallet game: Probability,
likelihood, and extended likelihood. \textit{The American Statistician}, 
\textbf{71}, 120-122.

\item  {Pawitan, Y. and Lee, Y. } (2021). Confidence as Likelihood. \textit{%
Statistical Science}, \textbf{36}, 509-517.

\item  {Pawitan, Y. and Lee, Y. } (2022). Rejoinder: Confidence as
Likelihood. \textit{Statistical Science}, \textbf{37}, 628-629.

\item  {Pawitan, Y. and Lee, Y. } (2024). \textit{Philosophies, Puzzles and
Paradoxes: A Statistician's Search for Truth}. (Boca Raton: CRC Press)

\item  {Popper, K. R. } (1959). New Appendices. (In \textit{The Logic of
Scientific Discovery}. New York: Routledge)

\item  {Popper, K. R. } (1972). \textit{Objective knowledge}. (Oxford:
Oxford University Press)

\item  {Ramsey, F. P. } (1926). Truth and Probability. (In R. B. Braithwaite
(Ed.), \textit{The Foundations of Mathematics and other Logical Essays}.
London: Routledge and Kegan Paul)

\item  Royal{, R. M. }(1997). \textit{Statistical Evidence: A Likelihood
Paradigm}, 2nd ed. (London : Chapman \& Hall)

\item  {Russell, B. } (1912). \textit{The Problems of Philosophy}. (New
York: Henry Holt and Co.)

\item  {Savage, L. J. } (1954). \textit{The Foundations of Statistics}. (New
York: Jon Wiley and Sons)

\item  {Savage, L. J. } (1961). The foundations of statistics reconsidered.
In \textit{Proceedings of the Fourth Berkeley Symposium on Mathematical
Statistics and Probability, Volume 1: Contributions to the Theory of
Statistics } (Vol. 4, pp. 575-587). (University of California Press)

\item  {Schweder, T. and Hjort, N. L. } (2016). \textit{Confidence,
Likelihood, Probability}. (Cambridge University Press)

\item  {Senn, S. } (2003). \textit{Dicing with Death: Chance, Risk and Health%
}. (Cambridge University Press)

\item  {Senn, S. } (2009). Comment on Harold Jeffreys's Theory of
Probability Revisited. \textit{Statistical Science}, \textbf{24}, 185-186.

\item  {Stein, C. } (1959). An example of wide discrepancy between fiducial
and confidence intervals. \textit{The Annals of Mathematical Statistics}, 
\textbf{30}, 877-880.

\item  {Welch, B. L. and Peers, H. W.} (1963). On formulae for confidence
points based on integrals of weighted likelihoods. \textit{Journal of the
Royal Statistical Society Series B: Statistical Methodology}, \textbf{25},
318-329.

\item  {Wilkinson, G. N.} (1977). On resolving the controversy in
statistical inference. \textit{Journal of the Royal Statistical Society
Series B: Statistical Methodology}, \textbf{39}, 119-144.

\item  {Von Mises, R. } (1928). \textit{Probability, Statistics and Truth},
2nd English ed. (Translated by H. Geiringer. London: Allen and Unwin)

\item  {Von Neumann, J. and Morgenstern, O. } (1947). \textit{Theory of
Games and Economic Behavior}, 2nd ed. (Princeton University Press)

\item  {Wagenmakers, E. J., Gronau, Q. F., Dablander, F. and Etz, A. }
(2022). The support interval. \textit{Erkenntnis}, \textbf{87}, 589-601.
\end{description}

\end{document}